\documentclass[aps,prb,twocolumn,groupedaddress]{revtex4-1}
\usepackage{amsmath}
\usepackage{graphicx}
\usepackage{color}

\begin{document}

\title{Charge transport in two dimensions limited by strong short-range scatterers: \\Going beyond parabolic dispersion and Born approximation.}

\author{B{\v r}etislav {\v S}op{\' i}k}
\affiliation{Central European Institute of Technology, Masaryk University, Kamenice 735, 62500 Brno, Czech Republic}

\author{Janik Kailasvuori}
\affiliation{International Institute of Physics, Universidade Federal do Rio Grande do Norte, 59078-400 Natal-RN, Brazil}
\affiliation{Max-Planck-Institut f\"ur Physik komplexer Systeme, N\"othnitzer Str. 38, 01187 Dresden, Germany}

\author{Maxim Trushin}
\affiliation{University of Konstanz, Fachbereich Physik, M703 D-78457 Konstanz, Germany}

\date{\today}

\begin{abstract}
We investigate the conductivity of charge carriers confined to a two-dimensional system with the non-parabolic dispersion $k^N$ with $N$ being an arbitrary natural number. A delta-shaped scattering potential is assumed as the major source of disorder. We employ the exact solution of the Lippmann-Schwinger equation to derive an analytical Boltzmann conductivity formula valid for an arbitrary scattering potential strength. The range of applicability of our analytical results is assessed by a numerical study based on the finite size Kubo formula. We find that for any $N>1$, the conductivity demonstrates a linear dependence on the carrier concentration in the limit of a strong scattering potential strength. This finding agrees with the conductivity measurements performed recently on chirally stacked multilayer graphene where the lowest two bands are non-parabolic and the adsorbed hydrocarbons might act as strong short-range scatterers.
\end{abstract}


\maketitle

\section{Introduction}
 
\begin{figure}[b]
\includegraphics[width=8cm]{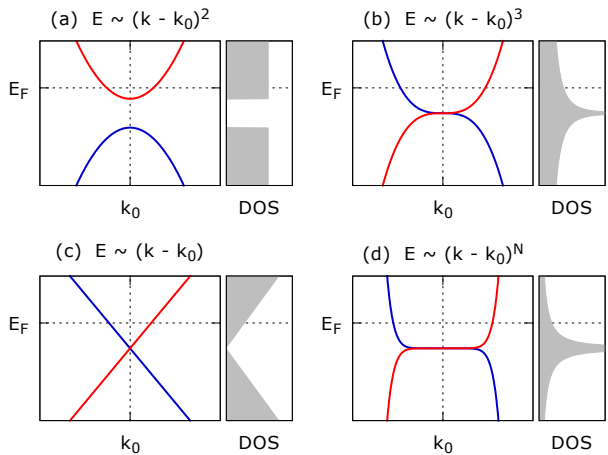}
\caption{\label{fig0} Panel (a) schematically shows a typical band structure for a two dimensional electron gas in III--V semiconductor heterostructures. The dispersion around the band minimum can be approximated by that of free electron with certain effective mass. If the crystal symmetry permits the contact between valence and conducting band, then the lowest order term in the Taylor expansion around $k=k_0$ can differ from the parabolic one. In particular (b) it can be cubic, as it is in the case of ABC stacked graphene \cite{PRB2008min}, (c) it can be linear, as it is for surface states in Bi$_2$Se$_3$ \cite{NatPhys2009zhang} or for single layer graphene \cite{PR1947wallace}, and (d) of arbitrary natural power $N$ which is the case of ABC-stacked multilayer graphene \cite{PRB2008min}. The short range disorder limited conductivity of the two-dimensional electron system with the $k^N$ dispersion is in the main focus of this paper.}
\end{figure}

The band theory provides a simple effective-mass description of charge carriers near the local minimum (maximum) of a conduction (valence) band almost in any semiconductor material where an energy gap separates the bands. Indeed, the energy dispersion can be expanded in the momentum near the bottom (top) of the conduction (valence) band. The linear term of the expansion is zero and the quadratic one mimics the free electron dispersion with the electron mass replaced by an effective mass.\cite{kittel1968} Already several decades ago it was pointed out,\cite{SPU1976berchenko} however, that the dispersion can differ from the parabolic one as long as the crystal symmetry permits the contact between the valence and conduction band in the quasi-momentum space. At the time, the problem was discussed in connection with mercury telluride which is a three-dimensional zero-gap semiconductor.\cite{SPU1976berchenko}  Recent advances in technology have made it possible to fabricate a few peculiar high-quality {\em two}-
dimensional gapless 
conductors: single layer graphene \cite{Nature2005novoselov,Nature2005zhang}, bilayer graphene \cite{NatPhys2006novoselov}, trilayer graphene\cite{NatNano2009craciun}, and topological insulators \cite{NatPhys2009xia,Science2009chen}, such as Bi$_2$Se$_3$ and Bi$_2$Te$_3$. The most of these materials demonstrate a carrier dispersion different from the parabolic one being standard for two-dimensional electron gases confined in III-V semiconductor heterostructures,\cite{datta1995} see Fig.~\ref{fig0}. In particular, this is the case of ABC-stacked N-layer-graphene \cite{PRB2008min,PRB2010zhang} or, equivalently, thin flakes of rhombohedral graphite \cite{CJP1958haering,Carbon1969mcclure}, where the charge carriers described by a simplified model with only nearest-neighbor interlayer hopping demonstrate a $k^N$ dispersion in both conduction and valence bands.

Electron transport in semiconductors is limited by the presence of localized impurities which can be described by either short-range or long-range potentials. The long-range potential represents charged impurities and can be approximated by a~screened Coulomb potential. The vacancies or some adsorbed molecules act as short-range scatterers which, in turn, can be approximated by a $\delta$-shaped potential. To derive the conductivity formula for the weak scattering potential one usually makes use of the semiclassical theory based on the Boltzmann equation with the golden-rule collision term.\cite{ashcroft1976} It is known the Fermi golden rule is derived within the first Born approximation. However, it is not always safe to say that the first Born approximation is valid as long as the potential strength is small. This is particularly important in the case of the short-range scatterers, as was mentioned in the very first chapter of the famous book by Peierls\cite{peierls1979}.

In this paper, we focus on the scattering by a short-range potential of carriers with non-parabolic dispersion. We investigate the applicability of the first Born approximation (weak scattering potential) and the resonant scattering approximation (strong scattering potential) and we also address breakdowns of these two complementary approaches. The problem of the Born approximation breakdown has recently arisen in the field of graphene, where the adsorbed hydrocarbons effectively act as strong short-range scatterers \cite{PRL2008robinson}. The phenomenon is known as scattering due to ``midgap states'' \cite{PRB2007stauber} or ``resonant scattering'' \cite{PRB2010yuan}. Since the problem has been reviewed by Peres in his colloquium paper \cite{RMP2010peres}, we do not adduce the complete list of references here. We refer to the recent publication by Ferreira \textit{et al.} \cite{PRB2011ferreira} where the Lippmann-Schwinger equation together with the $T$-matrix approach have been 
utilized to show that the strong short-range potential leads to a similar conductivity behavior in monolayer graphene with the linear dispersion and in bilayer with the parabolic bands. In what follows, we generalize this setting for particles with $k^N$-dispersion which is relevant, in particular, for the ABC-stacked N-layer-graphene.\cite{PRB2008min,PRB2010zhang}

First, we analytically solve the Lippmann-Schwinger equation for particles with a $k^N$-dispersion and subject to $\delta$-shaped scattering potential. Using our solution, we calculate the Boltzmann conductivity expression valid for any potential strength. Second, we compute the conductivity numerically utilizing the finite-size Kubo formula and compare the results of these two approaches. Third, we analyze the applicability of the first Born and resonant scattering approximations for different values of $N$. 

The main findings of this paper are: (i) The analytically derived formula for the conductivity in multilayer graphene ($N > 2$) reproduces numerical results very well in a broad range of conditions. (ii) We observe that the first Born approximation breaks down for fillings close to the neutrality point, and the transport in this region is described within the resonant scattering limit. This result is confirmed by the experimental evidence. (iii) At large enough filling the conductivity approaches the first Born approximation regime for any potential strength which is in contradiction with the monolayer graphene ($N = 1$), where under such conditions the Born approximation breaks down. This discrepancy is due to different asymptotic of the density of states in multilayer graphene. A comparison of the conductivity in the first Born approximation limit and the resonant scattering limit for different $N$ is done in Table~\ref{tab1}.

\section{Solution of the Lippmann-Schwinger equation and calculation of the conductivity}

Here, we utilize the effective low-energy two-band Hamiltonian for carriers in N-layer ABC-stacked graphene \cite{PRB2008min,PRB2010zhang} as a model system for particles with the $k^N$ dispersion. In the simplest case of negligible interlayer asymmetries and trigonal warping this Hamiltonian for a given valley can be parametrized as
\begin{equation}
\label{hN}
H_0= \gamma 
\left(\begin{array}{cc}
0 
& (k_x - {\rm i}  k_y)^N \\ 
(k_x + {\rm i}  k_y)^N & 0
\end{array} \right),
\end{equation}
where $k$ is the wave vector, and $\gamma$ is a constant depending on the hopping between sublattices. (Note that $\gamma$ and its dimension depend on $N$. In particular, note that $\gamma$ includes a factor $\hbar^N$.) We focus on the conduction band electrons which have the dispersion $E_k=\gamma k^N$, the density of states
\begin{equation} \label{dos}
D(k)=\frac{k^{2-N}}{2\pi \gamma N}, 
\end{equation}
and the eigenstates of the form
\begin{equation}
\label{ef}
 \phi_\mathbf{k}(\mathbf{r})=\frac{1}{\sqrt{2} L}
\left(\begin{array}{c}
1 \\ 
{\mathrm e}^{{\rm i}N\theta}
\end{array} \right) \exp({\rm i}{\bf k}\cdot{\bf r}),
\end{equation}
with $\theta=\mathrm{atan}(k_y/k_x)$. We consider the finite doping regime where $1 \ll l_{k_{\rm F}} k_{\rm F}$ applies, here $k_{\rm F}$ is the Fermi wave vector and $l_{k_{\rm F}}$ is the mean free path of such electron, and the Boltzmann approach is expected to be valid, in particular with the influence of the valence band being negligible. Thus, in contrast to our previous work, see Ref.~\onlinecite{JSMTE2013kailas}, it is the $k^N$ dispersion of carriers, rather than the chiral structure of the effective Hamiltonian, that is in the main focus of the present work. This approach is therefore not limited to graphene but can be applicable for the conductivity description of any other two-dimensional conductor with such a peculiar dispersion. 

To illustrate the practical application of the effective Hamiltonian we put it into a context of the ABC stacked trilayer graphene. Here, due to the split-off bands, the maximum quasiparticle energy in the considered two-band model is limited by the value of the order of $0.1\,\mathrm{eV}$.\cite{PRB2010zhang} The explicit expression for $\gamma$ in terms of the interlayer hopping parameter $t_\perp \simeq 0.4\,\mathrm{eV}$ and the characteristic velocity $v_0 \simeq 10^{8}\,\mathrm{cm/s}$ is given by $(\hbar v_0)^3 / t_\perp^2$.\cite{PRB2008min} The maximum carrier concentration thus may not exceed $\simeq 5 \times 10^{12}\,\mathrm{cm}^{-2}$ which is a value comparable with the one obtained from transport measurements in monolayer graphene. It is also worth to note that due to flatter bands for $N>2$ the density of states \eqref{dos} reaches higher values in the vicinity of the neutrality point, see Fig.~\ref{fig0}. This results into a stronger Thomas-Fermi screening of the charged 
impurities which makes considering the short-range disorder even more relevant.
  
As the scattering potential model, we utilize the $\delta$-shaped potential, $V(\mathbf{r})=V_0\delta(\mathbf{r})$. The total Hamiltonian with a single impurity reads $H=H_0+V(\mathbf{r})$. This model allows a non-perturbative analytical solution and results in an elegant conductivity formula valid for any potential strength $V_0$. To do that we follow the standard recipe used by Ferreira {\it et al.}, see Ref.~\onlinecite{PRB2011ferreira}, for the case of linear and parabolic bands with $N=1$, $2$.

\subsection{The Lippmann-Schwinger equation}

The Lippmann-Schwinger equation for the wave function $\psi_\mathbf{k}$ of a particle scattered on a single impurity reads
\begin{equation}
 \label{l-sch}
 \psi_\mathbf{k}(\mathbf{r})= \phi_\mathbf{k}(\mathbf{r}) + \int {\rm d}^2 \mathbf{r}' \, G_0 (\mathbf{r}-\mathbf{r}') V(\mathbf{r}') \psi_\mathbf{k}(\mathbf{r}'),
\end{equation}
where $G_0 (\mathbf{r}-\mathbf{r}')=\langle \mathbf{r} \vert (E+ i0 - H_0)^{-1} \vert \mathbf{r}' \rangle$ is the Green's function of the problem which can be written down explicitly as
\begin{equation}
 \label{green}
 G_0 (\mathbf{r}-\mathbf{r}') =(E + H_0)\int \frac{{\rm d}^2 \mathbf{k}'}{4 \pi^2} \frac{{\mathrm e}^{{\rm i}\mathbf{k}' \cdot (\mathbf{r}-\mathbf{r}')}}{(E + {\rm i}0)^2 -(\gamma k'^N)^{2}}.
\end{equation}
Since the scattering potential is $V(\mathbf{r}')=V_0 \delta(\mathbf{r}')$, the integral in Eq.~\eqref{l-sch} becomes trivial. The amplitude of the wave function at the origin $\psi_\mathbf{k}(0)$ is easy to calculate from the equation 
\begin{equation}\label{eq:scat-center-wave}
\psi_\mathbf{k}(0)= \phi_\mathbf{k}(0) + G_0 (0) V_0 \psi_\mathbf{k}(0)\,, 
\end{equation}
where $G_0(0)$ can be found straightforwardly from  Eq.~\eqref{green}, and for $N  > 1$ we obtain
\begin{equation}
 \label{G_0}
 G_0(0) = \frac{\pi}{2} D(k)\left[\mathrm{cot} (\pi/N)- {\rm i}\right]. 
\end{equation}
Thus, the Lippmann-Schwinger equation \eqref{l-sch} becomes rather simple and has the form
\begin{equation}
 \label{l-sch-2}
 \psi_\mathbf{k}(\mathbf{r})= \phi_\mathbf{k}(\mathbf{r}) +  \frac{ G_0 (\mathbf{r}) V_0 }
 {1-\frac{\pi}{2} D(k)\left[\mathrm{cot} (\pi/N) - {\rm i} \right] V_0} \phi_\mathbf{k}(0).
\end{equation}
The remaining task is to find $G_0(\mathbf{r})$. To do that we take the integral in Eq.~\eqref{green} in the polar coordinates $\{k',\theta' \}$. The integral over $\theta'$ results in the Bessel function of the first kind, and the subsequent integration over $k'$ gives a combination of Bessel functions and Meijer G-functions.\cite{prudnikov1990-3} To calculate the result of the action of $H_0$ on this expression in Eq.~\eqref{green}, $H_0$ should be also transformed into the polar coordinates. We do not express the general equation for $G_0 (\mathbf{r}-\mathbf{r}')$, since we employ its asymptotic form for $k\vert\mathbf{r}-\mathbf{r}'\vert\gg 1$ only, in which case the Green's function simplifies to
\begin{equation}
 \label{green-ass}
 G_0 (\mathbf{r}-\mathbf{r}') =-\sqrt{\frac{2}{\pi k |\mathbf{r}-\mathbf{r}'|}} {\mathrm e}^{{\rm i} k|\mathbf{r}-\mathbf{r}'|+ {\rm i}\frac{\pi}{4}}
  \frac{\pi}{2} D(k) (1+\sigma_\varphi),
\end{equation}
where the matrix $\sigma_\varphi$ is
\begin{equation}
 \label{sigma-matrix}
 \sigma_\varphi=
\left(\begin{array}{cc}
0 
& {\mathrm e}^{-{\rm i}\varphi N}  \\ 
{\mathrm e}^{{\rm i}\varphi N} & 0
\end{array} \right),
\end{equation}
with the angle $\varphi$ defined by a projection of a unit vector $(\mathbf{r} - \mathbf{r}') / \vert \mathbf{r} - \mathbf{r}'\vert = (\sin\varphi,\cos\varphi)^T$. Following Ref.~\onlinecite{PRB2011ferreira}, we approximate $|\mathbf{r}~-~\mathbf{r}'|\simeq\mathbf{r}- \mathbf{r}\cdot \mathbf{r}'/r$, identify the outgoing wave vector as $\mathbf{k}_\mathrm{out}=k \mathbf{r}/r$, and without loss of generality take the incident wave vector $\mathbf{k}$ along the $x$-axis. The wave function of the scattered particle can then be written as
\begin{equation}
 \label{psik}
 \psi_\mathbf{k}(\mathbf{r})= \phi_\mathbf{k}(\mathbf{r}) + f(\theta)\frac{{\mathrm e}^{{\rm i}kr}}{\sqrt{r}}\frac{1}{\sqrt{2} L}
\left(\begin{array}{c}
1 \\ 
{\mathrm e}^{{\rm i}N\theta}
\end{array} \right),
\end{equation}
where the scattering amplitude $f(\theta)$ reads
\begin{equation}
\label{f}
 f(\theta) = - \sqrt{\frac{2{\rm i}}{\pi k}} \frac{\frac{\pi}{2} D(k) V_0 \,\lbrack 1+ {\mathrm e}^{-{\rm i}N\theta} \rbrack}{1-\frac{\pi}{2} D(k) V_0 \left[\mathrm{cot}(\pi/N) - {\rm i} \right]},
\end{equation}
with $\theta=\angle (\mathbf{k},\mathbf{k}_\mathrm{out})$ being the scattering angle. Note the qualitative difference between the scattering amplitude~\eqref{f} for $N>2$ and the one derived in Ref.~\onlinecite{PRB2011ferreira} for the case of bilayer graphene $N=2$, where the denominator does not contain the term with $\cot (\pi / N)$.

\subsection{The Boltzmann dc conductivity}

To calculate the conductivity out of Eq.~\eqref{f} we need the total scattering cross section $\Sigma_T=\int {\rm d}\theta' (1-\cos\theta') |f(\theta')|^2 $. Calculating the integral we obtain
\begin{equation}
 \label{sigmaT}
 \Sigma_T = \frac{2\pi^2}{k} \frac{D^2(k) V_0^2}{\left[1-\frac{\pi}{2} D(k) V_0\, \mathrm{cot}(\pi/N)\right]^2 + \frac{\pi^2}{4}D^2(k) V_0^2}.
\end{equation}
The conductivity can be then written down in terms of either the momentum relaxation time $\tau_k^{-1}=n_i |\mathbf{v}_k| \Sigma_T$ or the mean free path $l_k=|\mathbf{v}_k| \tau_k$. (Here, $\mathbf{v}_k=\frac{\mathbf{k}}{\hbar}\gamma N k^{N-2}$ is the particle velocity, and $n_i$ is the concentration of scatterers.) In the latter case the conductivity is just given by
\begin{equation} \label{cond}
\sigma = \frac{e^2}{h} \frac{l_{k_{\rm F}} k_{\rm F}}{2}, 
\end{equation}
with $l_{k_{\rm F}}$ being the mean free path calculated for a given Fermi wave vector $k_{\rm F}$. 

The mean free path can be written explicitly as
\begin{equation}
\label{l}
l_{k_{\rm F}}=\frac{1}{n_i} \frac{k_{\rm F}}{2\pi^2}\frac{\left[1-\frac{\pi}{2} D(k_{\rm F}) V_0 \,\mathrm{cot}(\pi/N)\right]^2 + \frac{\pi^2}{4}D^2(k_{\rm F}) V_0^2}{D^2(k_{\rm F}) V_0^2}\,,
\end{equation}
and represents our main theoretical result. Let us discuss its limiting regimes, beginning with the Born approximation limit where $D(k_{\rm F})V_0 \ll 1$. In this limit we after expansion for small $D(k_{\rm F})V_0$ obtain
\begin{equation}
\label{born}
l^\mathrm{Born}_{k_{\rm F}} = \frac{1}{n_i} \frac{k_{\rm F}}{2\pi^2}
\frac{1}{D^2(k_{\rm F}) V_0^2},
\end{equation}
the first correction, $l_{k_{\rm F}} = l^\mathrm{Born}_{k_{\rm F}} + \Delta l^\mathrm{Born}_{k_{\rm F}}$, is of the order of $(D(k_{\rm F})V_0)^{-1}$,
\begin{equation}
\label{born-correction}
\Delta l^\mathrm{Born}_{k_{\rm F}} =  - \frac{1}{n_i}\frac{k_{\rm F}}{2\pi}\frac{\mathrm{cot}(\pi/N)}{D(k_{\rm F}) V_0},
\end{equation}
and the conductivity in the Born approximation reads
\begin{equation}
\label{conductivity-born}
\sigma^{\rm Born} = \frac{e^2}{h}\frac{n}{n_i}\frac{1}{\pi D^2(k_{\rm F}) V_0^2}\,.
\end{equation}
From \eqref{dos} we see that for $N > 2$ the density of states $D(k_{\rm F}) \to 0$ for $k_{\rm F} \to \infty$. This means that the Born approximation can be approached not only by decreasing the potential $V_0$ but also by increasing the filling $n$. On the other hand the density of states diverges as $k_{\rm F} \to 0$, so at this point the Born approximation breaks down. This is very different situation from $N = 2$ where the density of states $D$ is constant and $k_{\rm F}$ independent. Let us also note that the first correction term \eqref{born-correction} depends on $\cot(\pi/N)$ which is zero for $N = 2$ but increases with higher $N$ and makes the Born approximation limit less accessible.

The opposite limit of \eqref{l} is the regime where $1 \ll D(k_{\rm F})V_0$ which is also known as the regime of resonant scattering \cite{RMP2010peres}. We approach this limit for very strong potentials $V_0$ and also for $k_{\rm F} \to 0$ due to the divergence of $D(k_{\rm F})$. The mean free path can be in this regime expanded in powers of $1/(D(k_{\rm F})V_0)$. We obtain
\begin{equation}
\label{res}
l^\mathrm{res}_{k_{\rm F}} = \frac{1}{n_i} \frac{k_{\rm F}}{8} \left[1 +  \mathrm{cot}^2(\pi/N) \right],
\end{equation}
which does not depend on $D(k_{\rm F})V_0$. The first correction term, $l_{k_{\rm F}} = l^\mathrm{res}_{k_{\rm F}} + \Delta l^\mathrm{res}_{k_{\rm F}}$, is of the order of $(D(k_{\rm F}) V_0)^{-1}$,
\begin{equation}
\label{res-correction}
\Delta l^\mathrm{res}_{k_{\rm F}} =  - \frac{1}{n_i}\frac{k_{\rm F}}{2\pi}\frac{\mathrm{cot}(\pi/N)}{D(k_{\rm F}) V_0}.
\end{equation}
The conductivity as a function of carrier concentration turns out to be linear in $n$ for any $N$
\begin{equation}
 \label{conductivity-res}
 \sigma^\mathrm{res}=\frac{e^2}{h} \frac{n}{n_i}\frac{\pi}{4}\left[1+ \mathrm{cot}^2(\pi/N)\right].
\end{equation}
From this formula one can conclude that the resonant scattering regime has even more universal character than was found by Ferreira {\it et al.}\cite{PRB2011ferreira} The dependence of the conductivity on the carrier concentration is nearly linear at $V_0\to \infty$ not only in the case of $N=1$ or $N=2$ but for any other $N>2$. Similar to the Born approximation limit, the resonant scattering regime becomes less accessible for larger $N$.

Note, that all the formulas given above are valid for $N \geq 2$ only, and that the case of $N=1$ must be considered separately. As shown by Ferreira {\it et al}.,\cite{PRB2011ferreira} in order to calculate the integral over ${\bf k}'$ in Eq.~\eqref{green} it is necessary to introduce a momentum cut-off corresponding to the smallest length scale of the system, $R$. At $N = 1$ the Green's function thus reads
\begin{equation}
  \label{GreenN1}
  G_0(0) =  D(k)\left(\ln|kR|- {\rm i} \pi / 2 \right),
\end{equation}
which leads to the mean free path
\begin{equation}
\label{lN1}
l_{k_{\rm F}}=\frac{1}{n_i} \frac{k_{\rm F}}{\pi^2} \frac{\left[1 - D(k_{\rm F}) V_0 \,\ln|k_{\rm F}R|\right]^2 + \frac{\pi^2}{4}D^2(k_{\rm F}) V_0^2}{D^2(k_{\rm F}) V_0^2}\,.
\end{equation}
Here we can also distinguish two limiting regimes, the Born approximation regime, $l^{\rm Born}_{k_{\rm F}} = \tfrac{1}{n_i}\tfrac{k_{\rm F}}{\pi^2}(D(k_{\rm F})V_0)^{-2}$, and the resonant scattering regime, $l_{k_{\rm F}}^{\rm res} = \tfrac{1}{n_i}\tfrac{k_{\rm F}}{\pi^2}\left(\ln^2|k_{\rm F}R|+\tfrac{\pi^2}{4}\right)$. However, contrary to $N > 2$ for $N = 1$ the density of states $D(k_{\rm F})$ is zero for $k_{\rm F} = 0$ and diverges for $k_{\rm F} \to \infty$ which means that for fixed $V_0$ the conductivity approaches Born approximation regime for small filling $k_{\rm F} \to 0$ and the resonant scattering regime for $k_{\rm F} \to \infty$.

\begin{figure*}
\includegraphics[width=18cm]{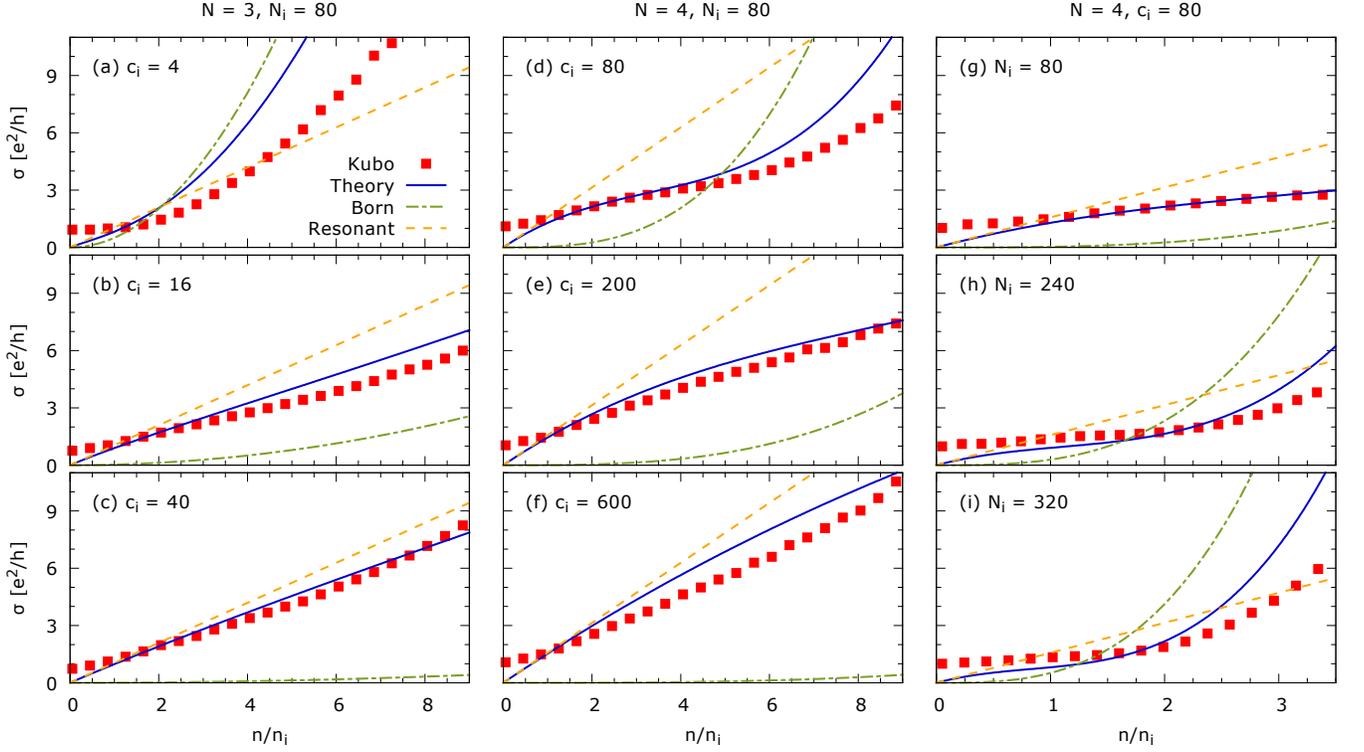}
\caption{\label{fig-1}The plots show the dependence of the conductivity on the concentration of carriers relative to the density of impurities $n / n_i$. The main observation is that at strong potential the conductivity becomes a linear function of $n/n_i$ for any $N>2$ which generalizes the conclusion made by Ferreira {\em et al.}\cite{PRB2011ferreira} for $N=1, 2$. Red squares represent the Kubo conductivity $\sigma^{\rm Kubo}$, blue solid line is the theoretical conductivity $\sigma$ calculated from  \eqref{cond} and \eqref{l} 
using the exact solution of the Lippmann-Schwinger equation, green dash-dotted line is the Born approximation limit $\sigma^{\rm Born}$ and yellow dashed line is the resonant scattering limit $\sigma^{\rm res}$. The strength of the potential $V_0$ is expressed by the dimensionless parameter $c_i$ using \eqref{V0_ci}. (a), (b), (c) Conductivity for $N = 3$, fixed number of impurities $N_i = 80$ and potential strengths $c_i = 4$, 16 and 40. (d), (e), (f) Conductivity for $N = 4$, $N_i = 80$ and potential 
strengths $c_i = 80$, $200$ and $600$. (g), (h), (i) Conductivity for $N = 4$, $c_i = 80$ and 
increasing number of impurities $N_i = 80$, 240, 320. Plots (h), (i) were calculated with $k_{\Lambda} = 20 \tfrac{2 \pi}{L}
$.
}
\end{figure*}

\begin{figure}
\includegraphics[width=8cm]{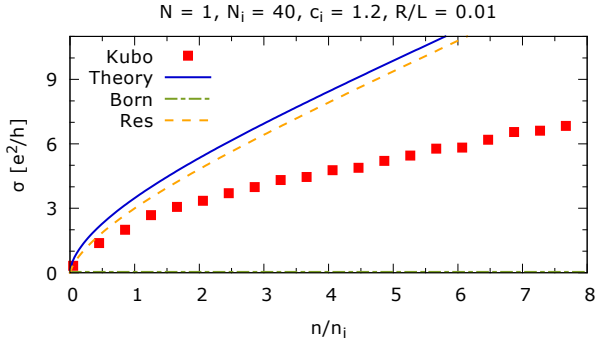}
\caption{\label{fig-2}The figure shows conductivity dependence on $n / n_i$ for $N = 1$, $N_i = 40$, $c_i = 1.2$ and $R / L = 0.01$. It reveals comparison of the Kubo conductivity $\sigma^{\rm Kubo}$ represented by red squares with the theory $\sigma$ from \eqref{lN1} and also its Born approximation limit $\sigma^{\rm Born}$ and resonant scattering limit $\sigma^{\rm res}$. We distinguish the lines in a same manner as in Fig.~\ref{fig-1}. The line $\sigma^{\rm Born}$ overlaps with zero conductivity axis. The resonant scattering limit $\sigma^{\rm res}$ was already derived by Ferreira {\it et al}., see Ref.~\onlinecite{PRB2011ferreira}.}
\end{figure}

\section{Numerical study of the finite-size Kubo formula}

In this section we compare the dc conductivity obtained analytically in the previous section from the Lippmann-Schwinger equation with results evaluated numerically from the finite-size Kubo formula. We begin with a discussion of details of the numerical method employed.

\subsection{Method}
The finite-size Kubo formula is given by
\begin{equation} \label{eq:cond-kubo}
 \sigma^{\rm Kubo}_{\lbrace {\bf R}_j \rbrace} = -{\rm i}\frac{\hbar e^2}{L^2}\sum_{n,n'} \frac{{\rm f}(E_n) - {\rm f}(E_{n'})}{E_{n} - E_{n'}}\frac{\langle n \vert {\bf v}_x \vert n' \rangle \langle n' \vert {\bf v}_x \vert n \rangle}{E_{n} - E_{n'} + {\rm i}\eta}\,.
\end{equation}
Here $L^2$ is a size of the system, ${\rm f}(E) = \Theta(E_{\rm F} - E)$ is the Fermi distribution function at zero temperature and $\eta = g_{\rm T} / (D(k_{\rm F})L^2)$ expresses broadening of levels due to the possibility of the particle to escape the system, with $g_{\rm T}$ being the dimensionless Thouless conductivity\cite{PRL2007nomura}. Vectors $\vert n \rangle$ and energies $E_n$ are eigenstates and eigenenergies of an effective mesoscopic Hamiltonian consisting of the kinetic term \eqref{hN} and potential term which is represented by $N_i = L^2 n_i$ scattering centers described by a $\delta$-shaped potential. The exact position of the scattering centers with respect to the underlying lattice is not addressed. We have
\begin{equation} \label{eq:hamilt-kubo}
 H(\lbrace {\bf R}_j \rbrace) = H_0 + \sum_{j = 1}^{N_i} V_0
\left(\begin{array}{cc}
1 & 0 \\ 
0 & 1
\end{array} \right)
\, \delta({\bf r} - {\bf R}_{j})\,,
\end{equation}
where locations $\lbrace {\bf R}_j \rbrace$ are randomly distributed in the continuum of the sample. For every distribution of the scattering centers $\lbrace {\bf R}_j \rbrace$ we diagonalize the Hamiltonian \eqref{eq:hamilt-kubo}, using a large momentum-space cut-off $k_{\Lambda}$, so that $k_{\rm F} < k_{\Lambda}$, and evaluate the conductivity \eqref{eq:cond-kubo}. This conductivity is then averaged over random distributions of the scattering centers, $\sigma^{\rm Kubo} = \langle \sigma^{\rm Kubo}_{\lbrace {\bf R}_j \rbrace} \rangle_{\rm av}$, until a sufficient precision is achieved. To improve the averaging we impose a small random shift $\delta{\bf k} \in ( -\tfrac{\pi }{L}, \tfrac{\pi }{L} ) \times ( -\tfrac{\pi }{L}, \tfrac{\pi }{L})$ on wave vector grid for every distribution $\lbrace {\bf R}_j \rbrace$ of scattering centers. Results in this work have been calculated using $g_{\rm T} = 12$, in agreement with the discussion in Ref.~\onlinecite{PRL2007nomura}. The momentum space cut-off $k_{\Lambda}$ 
was set to $16 \tfrac{2\pi}{L}$ if 
not stated differently. To express the strength of the potential $V_0$ we use the following parametrization 
\begin{equation} \label{V0_ci}
  V_0 = \gamma \left(\tfrac{2 \pi}{L}\right)^N L^2 \tfrac{N}{2\pi}\, c_i\,,
\end{equation}
with a dimensionless parameter $c_i$. This parametrization provides that both the kinetic and the potential term of the hamiltonian \eqref{eq:hamilt-kubo} scale like $(1/L)^N$ and so the Kubo formula \eqref{eq:cond-kubo} is independent on the length $L$ at zero temperature. That is because each of the two terms $\langle n \vert {\bf v}_x \vert n' \rangle / (E_n - E_n')$ scales like $\sim L$ which factors out $1 / L^2$ in front of the summation. Note that $L$ must be taken into account explicitly at finite temperatures, as it has been done for the description of thermally activated electron transport in gapped bilayer graphene.\cite{EPL2012trushin} For more details about the numerical method, see Refs.~\onlinecite{JSMTE2013kailas} and \onlinecite{PRL2007nomura}.

\subsection{Results}

The main results are depicted in Fig.~\ref{fig-1} which shows conductivity dependence on the filling relative to the density of impurities $n / n_i$ for several values of the power $N$ of the dispersion, number of impurities $N_i$ and potential strength $c_i$. Red squares represent the Kubo conductivity $\sigma^{\rm Kubo}$, the blue solid line is the theoretical conductivity $\sigma$ calculated from \eqref{cond} and \eqref{l}, the green dash-dotted line is the Born approximation limit $\sigma^{\rm Born}$ and the yellow dashed line is the resonant scattering limit $\sigma^{\rm res}$. The theoretical curves are plotted using $k_{\rm F} = \sqrt{4\pi n}$.

The most important observation in Fig.~\ref{fig-1}~(a)--(i) is that the theoretical conductivity $\sigma$ obtained from the Lippmann-Schwinger equation gives a good agreement with the Kubo formula conductivity $\sigma^{\rm Kubo}$ in a broad range of conditions. This is because we did not rely on the golden-rule relaxation time
employed in the previous papers \cite{PRB2011xu,PRB2010klos,JSMTE2013kailas}. Since the first correction term of $l_{k_{\rm F}}^{\rm Born}$ as well as $l_{k_{\rm F}}^{\rm res}$ is negative, the value of $\sigma$ is always below $\sigma^{\rm Born}$ or $\sigma^{\rm res}$. At the vicinity of the neutrality point the Kubo conductivity drops to a finite minimal conductivity which is given by interband scattering events. Since our theoretical approach considers intraband scattering exclusively, such contribution is not present in the theoretical conductivity and thus it goes to zero.

Plots (a), (b), (c) reveal the conductivity for $N = 3$, with fixed number of scattering centers $N_i = 80$ and increasing potential strength $c_i = 4$, 16 and 40. In (a) we see that both $\sigma$ and $\sigma^{\rm Kubo}$ follow $\sigma^{\rm Born}$ starting from very small values of $n/n_i$. As the potential strength $c_i$ grows in (b) and (c), the region where the resonant scattering limit is valid and $\sigma$ is linear in $n/n_i$, enlarges. Similar situation which confirms that this trend is universal for all $N > 2$, occurs in plots (d), (e), (f) for $N 
= 4$, $N_i = 80$ and potential strengths $c_i = 80$, $200$ and $600$. In (d) the conductivity $\sigma^{\rm Kubo}$ reaches the Born approximation behavior for large $n/n_i$. With higher values of $c_i$ in (e) and (f) both $\sigma$ and $\sigma^{\rm Kubo}$ get gradually closer to the linear behavior of the resonant scattering limit $\sigma^{\rm res}$ for all $n/n_i$ in the plot. The region of the Born approximation behavior is shifted to larger $n/n_i$ not shown here. Last column (g), (h), (i) shows the example with $N = 4$, constant $c_i = 80$ and increasing number of impurities $N_i = 80$, 240, 320. From \eqref{l} we see that $l_{k_{\rm F}}$ depends on the number of impurities only via factor $1 / n_i$ which scales the dependence of $\sigma$ on $n/n_i$. This is confirmed by numerical calculations. In (g) for $N_i = 80$ the conductivity $\sigma$ and $\sigma^{\rm Kubo}$ are very close to $\sigma^{\rm res}$ within the studied range of $n/n_i$. With increasing of $N_i$ in (h) and (i) the region, where $\sigma$ 
starts to follow the trend of Born approximation limit, gradually shifts to lower $n/n_i$. We conclude that $\sigma^{\rm Kubo}$ obeys the same scaling by $1/n_i$ as theoretical $\sigma$.

Fig.~\ref{fig-2} reveals a dependence of the conductivity on $n / n_i$ for $N = 1$, $N_i = 40$, $c_i = 1.2$ and represents a useful comparison with results from Fig.~\ref{fig-1}. We determine the smallest length scale as $R = 1 / k_{\Lambda}$ which gives $R / L = 0.01$. The theoretical conductivity overshoots the Kubo conductivity for all $n/n_i$, however, both conductivities obey a linear scaling for large $n/n_i$. The linear scaling of the conductivity in the diffusive regime was reported also by K\l{}os {\it et al}., see Ref.~\onlinecite{PRB2010klos}, that have done the numerical calculations within the Landauer approach.

The case of $N=2$ has been considered in Ref.~\onlinecite{PRB2010trushin} with the application to the pseudo-spin coherent conductivity of bilayer graphene. The first Born approximation \eqref{born} has been utilized there in order to fit the numerical Kubo conductivity curves. This approximation once established at $N=2$ remains valid for any $k_{\rm F}$ because the correction $(DV_0)^{-1}$ does not depend on $k_{\rm F}$ for the parabolic bands. This lucky circumstance made it possible to consider the pseudo-spin coherent terms in the Boltzmann equation within the golden-rule approximation and reach a good agreement between the numerical and analytical models even at lower carrier densities.\cite{PRB2010trushin} It is clear from Eq.~\eqref{born} now that this approach could not work for $N>2$ equally well as it did for $N=2$: The correction depends on $k_{\rm F}$ and the first Born approximation breaks down at low enough carrier concentrations.

\begin{table*}
{
\renewcommand{\arraystretch}{2}
\newcommand{\mc}[3]{\multicolumn{#1}{#2}{#3}}
\begin{tabular}{|l|c|l|c|l}\hline 
\mc{1}{|c|}{N} & \mc{2}{c|}{First Born approximation ($V_0\to 0 $)} & \mc{2}{c|}{Resonant scattering regime ($V_0\to \infty $)}\\\hline
× & $\sigma^{\rm Born}$ $[e^2/h]$& \mc{1}{c|}{$\delta^{\rm Born}$ $[\sigma^{\rm Born}]$} & $\sigma^{\rm res}$ $[e^2/h]$ & \mc{1}{c|}{$\delta^{\rm res}$ $[\sigma^{\rm res}]$}\\\hline
$N = 1$ & $\frac{2\gamma^2}{n_i V_0^2}$ & \mc{1}{c|}{ $-\frac{2V_0}{\gamma}\sqrt{\frac{n}{\pi}}\mathrm{ln}(R\sqrt{4\pi n})$} & $ \frac{n}{ n_i} \left[\frac{\pi}{2} + \frac{2}{\pi}\mathrm{ln}^2(R\sqrt{4\pi n})\right] $ & \mc{1}{c|}{$- \frac{2 \gamma}{V_0}\sqrt{\frac{\pi}{n}}\frac{\ln(R \sqrt{4 \pi n})}{\frac{\pi^2}{4} + \ln^2(R\sqrt{4 \pi n})}$}\\\hline
$N = 2$ & $\frac{16\pi \gamma^2}{n_i V_0^2} n$ & \mc{1}{c|}{$+\frac{V_0^2}{64\gamma^2}$} & $\frac{\pi}{4 n_i}n $ & \mc{1}{c|}{$+\frac{64\gamma^2}{V_0^2} $}\\\hline
$N > 2$ & $\frac{N^2 \gamma^2}{n_i V_0^2 }(4\pi n)^{N-1}$ & \mc{1}{c|}{$-\frac{V_0 \mathrm{cot}(\pi/N)}{2 N\gamma (4\pi n)^{\frac{N}{2}-1}} $} & $\frac{\pi n}{4n_i}\left[1+ \mathrm{cot}^2(\pi/N)\right]$  & \mc{1}{c|}{$-\frac{8N\gamma (4\pi n)^{\frac{N}{2}-1}\mathrm{cot}(\pi/N)}{V_0[1+ \mathrm{cot}^2(\pi/N)] }$}\\\hline
\end{tabular}
}
\caption{\label{tab1}Conductivity in the two limiting cases of the potential strength $V_0$ ---  the first Born approximation, $\sigma^{\rm Born}$, and the resonant scattering regime, $\sigma^{\rm res}$ --- and its first relative corrections $\delta^{\rm Born}$ and $\delta^{\rm res}$, with $\tilde \sigma^{\alpha} = \sigma^{\alpha} (1 + \delta^{\alpha})$. The rows for $N = 1$ and $N = 2$ are taken from Ref.~\onlinecite{PRB2011ferreira}, the third row follows from Eqs.~\eqref{born} and \eqref{res}. The dependence of conductivity on carrier concentration $n$ is qualitatively different for $N=1$, $N=2$, and $N>2$ in the Born approximation, but demonstrates universal linear response in the resonant scattering regime.}
\end{table*}

\section{Conclusions}

In this paper we have studied the dc transport of quasiparticles with $k^N$ dispersion in the presence of $\delta$-shaped scattering centers. This model can be applicable to multilayer ($N \geq 2$) graphene contaminated by hydrocarbons. Special attention was payed to two complementary limiting regimes --- the first Born approximation limit and the resonant scattering limit --- with respect to the limitations of these approaches. The results are summarized in Table~\ref{tab1} which shows $\sigma^{\rm Born}$, $\sigma^{\rm res}$ and their first relative corrections for $N = 1$, 2 and $N > 2$. We conclude that
\begin{itemize}
  \item Both the first Born approximation and the resonant scattering regime overestimate the conductivity for $N>2$, in contrast to the case of $N=2$, when the conductivity turns out to be underestimated.
  \item In contrast to the case of $N=2$, the conductivity correction to $\sigma^\mathrm{Born}$ is not quadratic but linear in $V_0$ for $N>2$. This makes the first Born approximation regime less accessible. The same is true for the resonant scattering regime with respect to $1 / V_0$.
  \item The conductivity corrections are concentration dependent for $N>2$. At large enough filling the conductivity approaches the Born approximation regime for any potential strength.
  \item In the limit of very strong scattering potential, $V_0\to \infty$, i.e., in the resonant scattering regime, the dependence of the conductivity on the carrier concentration is always linear for any $N > 2$. This generalizes the conclusion made by Ferreira {\em et al.}\cite{PRB2011ferreira} for $N=1, 2$.
\end{itemize}
These outcomes have been confirmed by the numerical conductivity calculation using the finite-size Kubo formula, see Figures~\ref{fig-1} and \ref{fig-2}.

Let us compare our theoretical results with experimental data for trilayer graphene, $N = 3$. Although related experimental studies are already present in the literature,\cite{NatNano2009craciun,NatPhys2011taychatanapat,NatPhys2011zhang,NatPhys2011lui,NatPhys2011bao} we found the comparison difficult. This is because there is often lack of information about the stacking (ABA or ABC) of the graphene sample on which the measurement was performed. For this reason we compare our results with work of Zhang {\it et al}.\cite{NatPhys2011zhang} only, since in this case we are sure the trilayer graphene with chiral ABC stacking was utilized, because in order to fit the data authors used Hamiltonian \eqref{hN} identical to the one discussed in this work. We observe that the conductivity in Figure~1~(e) of Ref.~\onlinecite{NatPhys2011zhang} is linear in $n$, similar to the conductivity in monolayer and bilayer.\cite{NatNano2009craciun} This suggests that the electron transport of the sample was in a regime of strong 
resonant scattering described by Eq.~(\ref{conductivity-res}) with $N=3$. We understand this as yet another evidence of significant impact of the scattering on short-range impurities on the electron transport in graphene.\cite{RMP2010peres}

\begin{acknowledgments}
M. T. thanks Wolfgang Belzig and Aires Ferreira for discussions and acknowledges financial support by the DFG through SPP 1285.
This work was also supported by the program "Employment of Newly Graduated Doctors of Science for Scientific Excellence" (CZ.1.07/2.3.00/30.0009) co-financed from European Social Fund and the state budget of the Czech Republic. The access to computing and storage facilities of the National Grid Infrastructure MetaCentrum provided under the program LM2010005 is also highly appreciated.
\end{acknowledgments}

\bibliography{graphene.bib,graphene2.bib,TI.bib}

\end{document}